\documentclass[a4,12pt]{article}
\usepackage{graphicx,psfrag,feynmp,amssymb,cite}

\renewcommand{\thefootnote}{\fnsymbol{footnote}}

\newcommand{\prepr}[1] {\begin{flushright}  {\bf #1} \end{flushright} \vskip 1.cm}
\newcommand{\titul}[1] {\boldmath \begin{center}{\Large {\bf #1 } } \end{center}
\vskip 0.8cm}

\newcommand{\autor}[1] {\begin{center}  {\bf \lineskip .3cm #1  }
                        \end{center} }

\newcommand{\lugar}[1] {\begin{center}  {\normalsize \bf \it #1   } \end{center}}
\newcounter{muni}

\begin{document}
\hbadness=10000
\pagenumbering{arabic}
\begin{titlepage}

\prepr{hep-ph/0209252\\
\hspace{30mm} KIAS-P02060 \\
\hspace{30mm} KEK-TH-845 \\
\hspace{30mm} September 2002}

\begin{center}
\boldmath
\titul{\bf Purely Leptonic $B$ decays at High Luminosity\\ 
$e^+e^-$ $B$ factories}
\unboldmath

\autor{A.G. Akeroyd$^{\mbox{1}}$\footnote{akeroyd@kias.re.kr},
S. Recksiegel$^{\mbox{2}}$\footnote{stefan@post.kek.jp} }
\lugar{ $^{1}$ Korea Institute for Advanced Study,
207-43 Cheongryangri-dong,\\ Dongdaemun-gu,
Seoul 130-012, Republic of Korea}

\lugar{ $^{2}$ Theory Group, KEK, Tsukuba, Ibaraki 305-0801, Japan }

\end{center}

\vskip2.0cm

\begin{abstract}
\noindent{High Luminosity upgrades of the KEK--B collider are
being discussed. We consider the role of the purely 
leptonic decays $B^\pm \to l^\pm \nu$ and $B^0 \to l^+l^-$
in motivating such an upgrade. These decays are very sensitive
to $R$ parity violating extensions of the MSSM, and
we show that future runs of the KEK--B factory can be
competitive with high energy colliders for probing such models.}

\end{abstract}

\vskip1.0cm
\vskip1.0cm
\boldmath
\unboldmath

\vskip3.0cm
\begin{center}
\noindent
(Talk presented at the Third Workshop on a High Luminosity B Factory,\\
6-7 August, 2002, Shonan Village, Kanagawa, Japan) 
\end{center}

\end{titlepage}
\thispagestyle{empty}
\newpage

\pagestyle{plain}
\thispagestyle{plain}
\renewcommand{\thefootnote}{\arabic{footnote} }
\setcounter{footnote}{0}

\section{Introduction}
High luminosity upgrades of the KEK--B collider (hereafter to be
called Super KEK--B) along with corresponding
upgrades of the BELLE detector are being actively discussed
\cite{SuperKEKB:2002bb}. Similar plans are also envisaged 
for BABAR \cite{:2001xw}.
Luminosities of $10^{35}$cm$^{-2}$s$^{-1}$ (or greater) are 
deemed attainable by the year 2006/2007, which would enable data
samples of a few $10^9$ $B$ mesons per year of operation. 
Convincing theoretical motivation and proven complementarity
to hadronic $B$ factories are crucial in order to justify such an upgrade. 
In this talk we focus on the purely leptonic decays of $B$ mesons, 
$B^\pm_{u,c} \to l^\pm\nu$ and $B^0_{d,s}\to l^+l^-$, which remain elusive
to date, and discuss their role in motivating such a high luminosity 
option. We stress the theoretical interest in searching for 
these 18 decay modes. $B^0_s$ and $B^\pm_c$ mesons are too heavy to be 
produced at the
$\Upsilon(4S)$ but their decays can be probed at hadronic $B$ factory
experiments like LHC--B and B--TeV. These experiments
will be operating at the time of the proposed Super KEK--B. 
Super KEK--B would be an ideal
environment to search for $B^\pm_u\to l^\pm\nu$ and $B^0_d\to l^+l^-$
and has powerful advantages over the hadronic $B$ factories
in several channels. Together these two distinct $B$ factories
can provide much improved coverage of these 18 decays, 
which would offer measurements of the decay constants and/or probe models
beyond the SM, especially $R$ parity violating extensions of the MSSM.

\boldmath
\section{Leptonic $B$ meson decays.}
\unboldmath

\begin{figure} \begin{center}
\includegraphics[width=7cm]{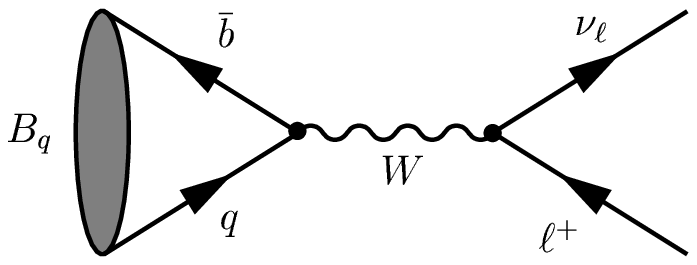}
\caption{Leptonic $B^\pm$ decay in the Standard Model: $W$ exchange
\label{wexchange}}
\end{center} \end{figure}

\suppressfloats[t] 
Purely leptonic $B$ decays form a particularly appealing group of
rare decays. For $B^\pm_{u,c}$ they proceed via annihilation 
to a $W^\pm$ in the SM (see Fig.~\ref{wexchange}), and due to 
helicity suppression, the rate is proportional to $m^2_l$.
Observation of such decays would allow a direct measurement of the 
$B^\pm$ meson decay constants.
The tree--level partial width is given by (where $q=u$ or $c$):
\begin{equation}
\Gamma(B^+_q\to \ell^+\nu_\ell)={G_F^2 m_{B_q} m_l^2 f_{B_q}^2\over 8\pi}
|V_{qb}|^2 \left(1-{m_l^2\over m^2_{B_q}}\right)^2
\end{equation}
The decay constants are accurately known for $\pi^\pm$ and $K^\pm$, while
those for $D^\pm$ mesons will be measured well at CLEO--c and 
(with probably less precision) at current runs of the $B$ factories. For
$B^\pm_u\to l^\pm\nu$ the severe suppression factor of
$|V_{ub}|^2$ renders the rates smaller than the analogous decays 
of the lighter mesons. The branching ratios (BRs)
for $B^\pm_c\to l^\pm\nu$
are larger by a factor $|V_{cb}/V_{ub}|^2$ but this is compensated
by the suppression in the production cross-section 
relative to that of $B^\pm_u$. The BRs for $B^\pm_{u,c}\to l^\pm\nu$
in the SM and the current experimental
limits are given in Table 1. Note that the LEP search for
$B^\pm_u\to \tau^\pm\nu_{\tau}$ is also sensitive to
$B^\pm_c\to \tau^\pm\nu_{\tau}$, which has the same experimental
signature. Thus the displayed limit constrains the BR of an admixture
of $B_u^\pm$ and $B_c^\pm$. BABAR studies have shown 
that 17 (8) events could be expected for $B^\pm_u\to \tau^\pm\nu$
($B^\pm_u\to \mu^\pm\nu$) with 500 
fb$^{-1}$ \cite{:2001xw}. This is clearly insufficient to make a
serious measurement of the decay constant $f_{B_u}$. Data samples
an order of magnitude larger, which are feasible at
Super KEK--B, would enable the first direct measurements of $f_{B_u}$. 
\begin{table}\begin{center}
\begin{tabular} {|c|c|c|c|c|} \hline
Decay & SM Prediction & CLEO   & {\sc Belle} & LEP / Tevatron \\ \hline
 $B_u^+\to e^+\nu_e$ & $9.2\times 10^{-12}$ & $\le 1.5\times 10^{-5}$
 \cite{Artuso:1995ar}& 
 $\le 5.4\times 10^{-6}$\cite{BELLE:2001} & $\otimes$ \\ \hline
 $B_u^+\to \mu^+\nu_{\mu}$ & $3.9\times 10^{-7}$ & 
$\le 2.1\times 10^{-5}$ \cite{Artuso:1995ar}
  & $\le 6.8\times 10^{-6}$ \cite{BELLE:2001} & $\otimes$ \\ \hline
  $B_u^+\to \tau^+\nu_{\tau}$ & $8.7\times 10^{-5}$ 
& $\le 8.4\times 10^{-4}$ \cite{Browder:2000qr} & $\otimes$ 
& $\le 5.7\times 10^{-4}$ \cite{Acciarri:1996bv} / $\otimes$  \\ \hline
 $B_c^+\to e^+\nu_e$ & $2.5\times 10^{-9}$ & $\times$  
& $\times$ & $\otimes$ \\ \hline
 $B_c^+\to \mu^+\nu_{\mu}$ & $1.1\times 10^{-4}$  
 & $\times$ & $\times$ & $\otimes$ \\ \hline
 $B_c^+\to \tau^+\nu_{\tau} $ & $2.6\times 10^{-2}$ 
& $\times$ & $\times$ & (see text) / $\otimes$  \\ \hline
\end{tabular}\end{center}
\caption{SM predictions and current experimental limits from various 
machines. \label{explimits}}
\end{table}

The decays $B^0_{d,s}\to l^+_il^-_j$, for 
$i=j$ proceed via higher order diagrams in the SM
and are very suppressed. For  $i\ne j$ they are forbidden in the minimal
SM. Thus measurements of the decay constants $f_{B_d}$ and $f_{B_s}$
are not feasible using these channels.
However, the leptonic $B^0$  decays can play
an additional role of probing models beyond the SM. Due to their 
very small rates in the SM, observation of these decays 
would be unequivocal evidence of new physics. Although the BRs
for $B^\pm\to l^\pm\nu$ are larger, these decays can also
probe physics beyond the SM. The
MSSM with $R$ parity conservation can sizeably enhance at most a 
select few of these decays, i.e.
i) $B^\pm_u\to \tau^\pm\nu$ via tree level exchange of $H^\pm$ to 
current experimental limits \cite{Hou:1992sy}, 
ii) $B^\pm_u\to \mu^\pm\nu$ to the sensitivity of the larger 
data samples at the B factories\cite{Hou:1992sy}, 
iii) $B^0_s \to \mu^+\mu^-$ can also be enhanced by neutral Higgs
penguin diagrams \cite{Dedes:2001fv}, and could be in range at 
the Tevatron Run II. All
other decays are too suppressed to be in reach in this model
at current and planned experiments. 

In contrast, such decays are very sensitive to $R$ parity violating 
interactions of the scalar sparticles. 
\cite{Guetta:1997fw,Baek:1999ch,Dreiner:2001kc,Akeroyd:2002cs,
Jang:1997jy}. In such models squarks and sleptons 
may mediate all 18 decays at tree-level, and
thus these decays may play a crucial role in constraining or confirming
$R$ parity violating extensions of the MSSM.

\boldmath
\section{$R$ parity violation}
\unboldmath
The main motivation for $R$ parity violating SUSY 
\cite{Dreiner:1997uz,Allanach:1999ic}
is to account for the observed neutrino oscillations without
increasing the particle content of the MSSM \cite{Aulakh:1982yn}.
Although there are other mechanisms which can provide neutrino mass
(see \cite{Altarelli:2002hx} for a recent review) 
$R$ parity violation should be taken as a 
reasonable candidate. At upcoming high energy
colliders it is important to find distinctive signatures of models which
can generate neutrino mass. The large number of couplings in 
the most general $R$ parity violating models, which is sometimes seen 
as a deficiency of the model, does bestow it a rich phenomenology. 
Thus if $R$ parity violation is the
mechanism of neutrino mass generation, one would expect it
to provide many other phenomenological signatures. This is in contrast
to other mechanisms of neutrino mass generation which sometimes offer 
relatively few low energy signatures. 

The superpotential is given by:
\begin{equation} \label{potential}
W_{R}={1\over 2}\lambda_{ijk}L_iL_jE^c_k+
\lambda'_{ijk}L_iQ_jD^c_k+{1\over 2}\lambda''_{ijk}U^c_iD^c_jD^c_k
\end{equation}
Bilinear terms $\mu_iL_iH_2$ are also possible, but have 
negligible impact on the annihilation decays we consider. 
Since the $\lambda''_{ijk}U^c_iD^c_jD^c_k$ term
can mediate proton decay it is customary to assume that 
the $\lambda''$ couplings vanish due to some discrete symmetry 
(e.g. baryon parity).

The simplest approach to $R$ parity violating phenomenology is
to assume that a single $R$ parity violating coupling
in the {\sl weak basis} $(\lambda'_{ijk}$)
is dominant with all others negligibly small. 
It was shown that such an approach leads to several non-zero 
$R$ parity violating couplings in the {\sl mass basis} 
$(\bar\lambda'_{imn})$ due to quark mixing \cite{Agashe:1995qm}:
\begin{equation}
\bar\lambda'_{imn}=\lambda'_{ijk}V^{\rm KM}_{jm}\delta_{kn}
\end{equation}
Here we have assumed that all quark mixing lies in the up--type sector,
so that the mixing matrix is the usual Kobayashi--Maskawa matrix $V^{\rm KM}$.
This simplification avoids the appearance of the right--handed quark
mixing matrix and gives the most conservative limits
on the $R$ parity violating couplings, which would otherwise
be constrained more severely from the decay $K^\pm\to \pi^\pm\nu
\overline\nu$. A realistic $R$ parity violating model would have many
non--zero couplings {\sl in the weak basis} and so 
in general would have a very rich phenomenology
provided the couplings are not too small.

$R$ parity violating trilinear couplings generate a neutrino mass at 
the 1--loop level. To fit the observed neutrino
oscillations, values of $10^{-3}\to 10^{-4}$ are required,
although there is an additional dependence on the L--R mixing in the 
squark/slepton sector which complicates the exact bounds which can be 
derived. The recently calculated two loop contributions 
can also be important \cite{Borzumati:2002bf}.
In purely bilinear models the sneutrino vacuum expectation values
provide a mass at tree--level to one neutrino, while
all three neutrinos receive a 1--loop mass from the 
effective $\lambda'_{i33}$ couplings. The currently favoured large 
mixing angle solution can be accommodated with just trilinear
couplings, or in purely bilinear models if 
the universality condition is relaxed 
\cite{Kaplan:1999ds,Hirsch:2000ef,Chun:2002rh}.
Purely bilinear models, while having the advantage of possessing
fewer free parameters, are not expected to give rise to enhanced 
leptonic decays due to the strong constraints on $\mu_i$ from neutrino
oscillations. 

\boldmath
\section{Distinctive signatures of $R$ parity violation}
\unboldmath
We first briefly list some distinctive signatures of
$R$ parity violation.

\subsection{Single production of sparticles}
Single production of sparticles \cite{Dreiner:pe}
can be observable at future high--energy
colliders if the relevant $R$ parity violating
coupling is ${\cal O} (0.1)$. Several $\lambda,\lambda'$
couplings are relatively weakly constrained and may be of this order. 
Examples of single production at hadron colliders 
include slepton strahhlung \cite{Borzumati:1999th} and slepton production as 
a resonance in the $s$--channel \cite{Kalinowski:1997zt}.

\subsection{Decaying neutral or charged LSP}
Perhaps the most robust test of $R$ parity violation. While
the stable (necessarily neutral) LSP in $R$ parity conserving models 
provides a missing energy signature at colliders, 
in $R$ parity violating models it decays, usually into visible particles
\cite{Dawson:1985vr}. The decays of a LSP neutralino \cite{Hirsch:2000ef,
Chun:2002rh,Borzumati:2001pw}
and sneutrino \cite{Chun:2001mm} have been investigated.
Since the LSP is unstable it may be charged, and the case of a
LSP $\tilde\tau^\pm$ has been considered in \cite{Akeroyd:2001pm}.
If $\lambda,\lambda'> {\cal O}(10^{-5})$, then the LSP decays in the detector.

\subsection{Low energy probes}
Many low energy processes are sensitive to $R$ parity violating couplings
and an extensive literature exists on the various
constraints which can be derived e.g. see \cite{Bhattacharyya:1996nj}.
Recent single bounds on all the $R$ parity 
violating couplings are listed in \cite{Allanach:1999ic}, 
assuming that one coupling is dominant. Bounds on products of couplings which
mediate a given process are sometimes stronger than the product of the
bounds on the individual couplings.

\boldmath
\section{$B$ decays as probes of $R$ parity violation}
\unboldmath
%
\boldmath
\subsection{The decays $B^\pm\to l^\pm\nu$ and $B^0\to l^+_il^-_j$}
\begin{figure} \begin{center}
\includegraphics[width=11cm]{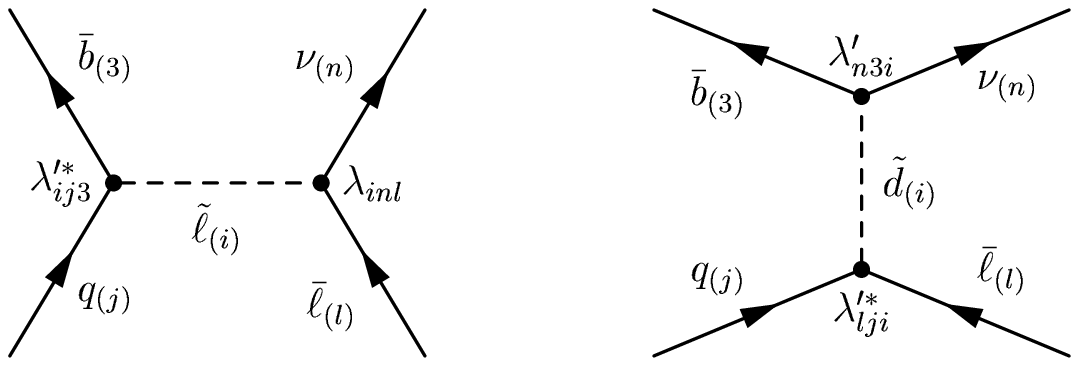}
\caption{Leptonic $B^\pm$ decay in $R$ parity violating models: 
sparticle exchange
\label{sparticleexchange}}
\end{center} \end{figure}

\unboldmath
\suppressfloats[t]
The purely leptonic decays $B^\pm_{u,c}\to l^\pm\nu$  
\cite{Baek:1999ch,Dreiner:2001kc,Akeroyd:2002cs}
are sensitive at tree level to 
$R$ parity violating trilinear interactions, 
and thus these decays constitute excellent probes of
the model. The relevant Feynman diagrams are depicted in 
Fig.~\ref{sparticleexchange}
and consist of $s$-- and $t$--channel exchange of sparticles. 
These additional channels modify the SM rate (Eq.1) by
\begin{equation}
m_l \quad\to\quad \left(1+{\cal A}^q_{ln}\right)m_l - 
(R_l+ {\cal B}^q_{ln})M_{D_q}\,,
\end{equation}
Here $R_\ell=m_\ell M_{D_q} \tan^2\beta/M_{H^\pm}^2$ 
stems from $R$ parity conserving SUSY charged Higgs exchange which
we will not consider further. This contribution is most 
important for the $B^\pm_{u,c}\to \tau^\pm\nu_\tau$. 
The ${\cal A}({\cal B})$ coupling corresponds to squark (slepton) exchange.
\begin{eqnarray}
{\cal A}^q_{ln}&=&{\sqrt 2\over 4G_FV_{qb}}\sum_{i,j=1}^3{1\over 
2 m^2_{\tilde q_i}}V_{qj}\lambda'_{n3i}\lambda'^*_{lji},\\ 
{\cal B}^q_{ln}&=&{\sqrt 2\over 4G_FV_{qb}}\sum_{i,j=1}^3{2\over 
m^2_{\tilde \ell_i}}V_{qj}\lambda_{inl}\lambda'^*_{ij3},
\end{eqnarray} 
$q=u,c$. These formulae were derived in \cite{Baek:1999ch}.
The analogous expressions for $B^0_{d,s}\to l^+l^-$ can be
found in \cite{Jang:1997jy}. The dominant ${\cal B}^q_{ln}$ term
(which is not helicity suppressed) requires one non--zero $\lambda$ 
{\it and} one non--zero $\lambda'$. For a given decay 
$B^\pm\to l^\pm\nu$ (i.e. fixing $l$ and $n$) there are nine
different combinations ($i,j=1,2,3$) of $\lambda\lambda'$ which contribute.
For a given $B^0_{d,s}\to l^+_il^-_j$ there are 6 combinations. 
Note that in practice the flavour of the neutrino ($n$) cannot be 
identified and so one must also sum over $n$.

Therefore it
is clear that these leptonic decays are sensitive to many 
combinations of the $R$ parity violating couplings. 
All 18 decays can be enhanced by the ${\cal B}$ coupling alone up to current
experimental sensitivity. Thus they might be observed any time 
in the current runs of the $B$ factories, and every new search 
improves limits on the relevant ${\cal B}$ coupling, which can be 
transformed into limits on 
particular combinations of $\lambda\lambda'$. Note that in a general
$R$ parity violating model one would expect many non--zero 
$\lambda\lambda'$ combinations
which would cause constructive and/or destructive interference in the 
${\cal B}$ coupling. If one or more of these decays were observed in current 
runs then there would be strong motivation for Super KEK--B to measure
more precisely the BRs and to search for other purely leptonic decays
which might be similarly enhanced. Super KEK--B would be ideal for 
these measurements.

If no purely leptonic decays are observed in the current runs, 
except for a few events
for $B^\pm_u\to \tau^\pm\nu_{\tau}$ and  $B^\pm_u\to \mu^\pm\nu_{\mu}$ 
consistent
with the SM prediction (see section 2), then the limits on the
${\cal B}$ coupling would be ${\cal O} (10^{-5})$. This would improve
to ${\cal O} (10^{-6})$ at Super KEK--B.

\subsection{Comparison with Hadronic machines}

At the proposed time of operation of Super KEK--B (circa 2006/2007)
the hadronic experiments LHC--B and B--TeV are expected to be
running. These latter experiments aim to accumulate
${\cal O} (10^{11-12})$ $B$ mesons, which is more than what is
anticipated at Super KEK--B ${\cal O} (10^{9-10})$.
In Table 2 we list the 18 decays and compare the potential of 
the hadronic machines and Super KEK--B to carry out searches.

\begin{center}
\begin{tabular} {|c|c|c|} \hline
 & Hadronic $B$ factory & KEK--B    \\  \hline
$B^+_c\to e^+\nu,\mu^+\nu,\tau^+\nu$  & maybe &  no \\  \hline 
$B^+_u\to e^+\nu,\mu^+\nu,\tau^+\nu$  & maybe & 
 yes \\  \hline
$B^0_s\to e^+e^-,e^+\mu^-,\mu^+\mu^-$ &  yes 
& no  \\  \hline
 $B^0_s\to e^+\tau^-,\mu^+\tau^-,\tau^+\tau^-$ 
& maybe & no    \\  \hline
 $B^0_d\to e^+e^-,e^+\mu^-,\mu^+\mu^-$ & yes
 &  yes \\  \hline
 $B^0_d\to e^+\tau^-,\mu^+\tau^-,\tau^+\tau^-$ 
& maybe &  yes   \\  \hline
\end{tabular}
\end{center}

One can see that Super KEK--B is superior at searching for 
decays with missing energy or at least one 
$\tau^\pm$. These channels are much more problematic for hadronic colliders,
which up to now (e.g. the Tevatron) have only concentrated on searching
for bi--lepton events. The capability of a hadronic machine to search 
for these missing energy and/or $\tau^\pm$ channels merits
further consideration in our opinion.
Hence Super KEK--B extends the coverage from 6 channels to 12, and
in addition would offer complementary measurements of 
$B^0_d\to e^+e^-,e^\pm\mu^\mp,\mu^+\mu^-$. 
 
\begin{figure} \begin{center}
\includegraphics[width=16cm]{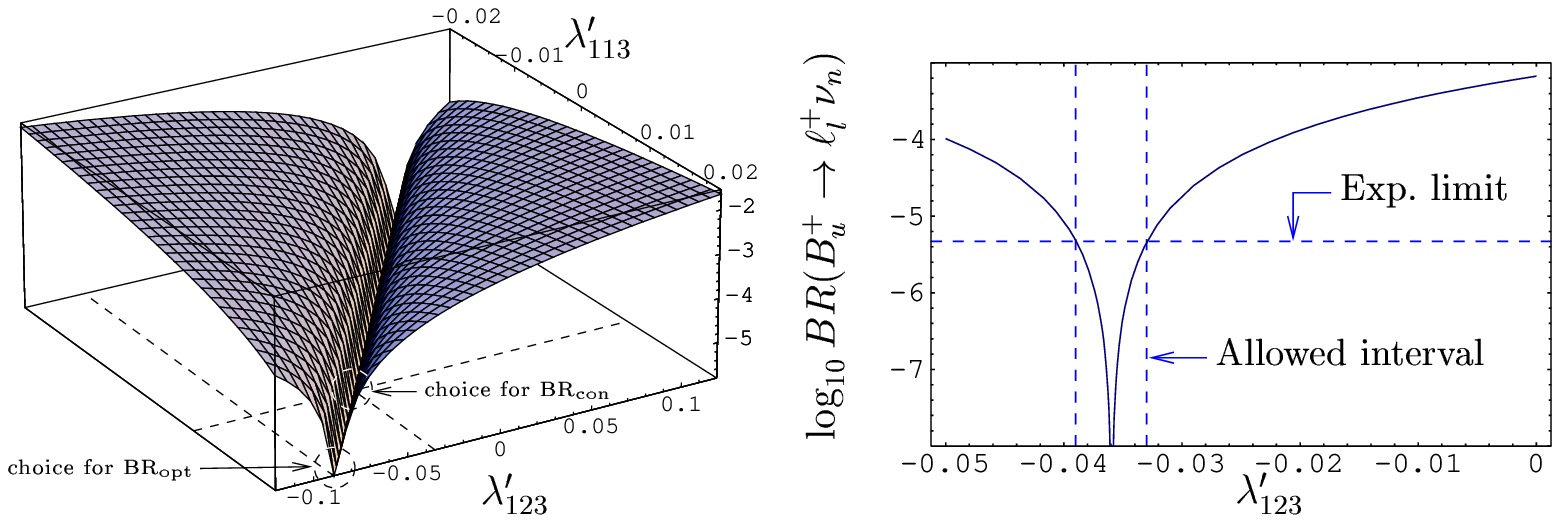}
\caption{Destructive interference in the $R$ parity violating contributions to
$B^\pm_u\to l^\pm\nu$ allowing large BRs for $B^\pm_c\to l^\pm\nu$ while
respecting experimental bounds on $B^\pm_u\to l^\pm\nu$ 
\label{valley}}
\end{center} \end{figure}

Note that only the hadronic machines
could search for $B^+_c\to e^+\nu,\mu^+\nu$, 
for which no experimental limits exist. We showed in \cite{Akeroyd:2002cs}
that these BRs can be greatly enhanced to ${\cal O}(10^{-2})$.
The ratio $BR(B^\pm_u\to l^\pm\nu)/BR(B^\pm_c\to l^\pm\nu$) is of 
theoretical interest since these two BRs would be correlated in the simplest
$R$ parity violating models with a single dominant $\lambda\lambda'$
combination.  This correlation would be relaxed if more combinations also
contribute non--negligibly to ${\cal B}$. In fact, with only one 
non--zero $\lambda$
and two non--zero $\lambda'$ couplings there can be destructive interference
that suppresses BR($B^\pm_u\to l^\pm\nu$) safely below experimental limits
while allowing for ${\cal O}(10^{-2})$ for BR($B^\pm_c\to l^\pm\nu$).
A diagram illustrating the destructive interference is shown 
in Fig.~\ref{valley}, 
taken from \cite{Akeroyd:2002cs} where the difference between our 
conservative (``con'') and optimistic (``opt'') estimates is explained.
BRs for $B^\pm_c\to l^\pm\nu$ of ${\cal O}(10^{-2})$ are attainable in
both cases.

For the reasons shown above, the measurement of the ratio 
$BR(B^\pm_u\to l^\pm\nu)/BR(B^\pm_c\to l^\pm\nu$) is an example
of how hadronic $B$ factories and Super KEK-B together could shed much light
on the underlying structure of $R$ parity violating models and other
models of physics beyond the SM. Studying correlations among 
rare $B$ decays in $R$ parity violating models has also been emphasized in
\cite{Saha:2002kt}.

\boldmath
\section{Conclusions}
\unboldmath
We have discussed the role of the purely leptonic decays of the
$B$ mesons in motivating a High Luminosity upgrade of the KEK--B collider.
We showed that such decays are very sensitive to $R$ parity violating
extensions of the MSSM and offer alternative ways of constraining/confirming
$R$ violating theories which are competitive with other high energy colliders.

\boldmath
\section*{Acknowledgements} 
\unboldmath

We thank the organizers of the Third Workshop on a Higher Luminosity
B Factory. We wish to thank F. Borzumati for useful discussions and comments.
S.R.\ was supported by the Japan Society for the Promotion of Science (JSPS).

\renewcommand{\theequation}{B.\arabic{equation}}
\setcounter{equation}{0}

\end{document}